\documentclass[amsmath,nofootinbib,amssymb,prl,twocolumn]{revtex4}
\pdfoutput=1

\usepackage[utf8]{inputenc}
\usepackage{float}
\usepackage{epsfig}
\usepackage{graphicx}
\usepackage{color}
\usepackage{tabularx}
\usepackage{color}
\usepackage{verbatim}
\usepackage[inline, final]{showlabels}
\usepackage{tensor}
\usepackage{enumerate}
\usepackage{subfigure}

\usepackage{color}
\usepackage[bookmarks,colorlinks,linkcolor=webblue,citecolor=webgreen]{hyperref}	
\definecolor{webgreen}{rgb}{0, 0.5, 0} 
\definecolor{webblue}{rgb}{0, 0, 0.5} 
\definecolor{webred}{rgb}{0.5, 0, 0} 

\newcommand{\av}[1]{\left\langle #1 \right\rangle}
\newcommand{\Od}[1]{\mathcal{O}\!\left( #1 \right)}
 
\newcommand{\ii}{\text{i}}
\newcommand{\di}{\text{d}}

\renewcommand{\v}[1]{\boldsymbol{#1}}

\renewcommand{\d}[2]{\frac{\text{d} #1}{\text{d} #2}}

\begin{document}
\preprint{}
\title{Magnetic fields from cosmological bulk flows}
\author{J.\,A.\,R.\,Cembranos\footnote{cembra@ucm.es}, 
        A.\,L.\,Maroto\footnote{maroto@ucm.es},
        and H.\,Villarrubia-Rojo\footnote{hectorvi@ucm.es}}
\affiliation{Departamento de  F\'{\i}sica Te\'orica and Instituto de Física de Partículas y 
			 del Cosmos IPARCOS, Universidad Complutense de Madrid, E-28040 Madrid, Spain}
\date{\today}
\begin{abstract}
    We explore the possibility that matter bulk flows could generate the required vorticity 
    in the electron-proton-photon plasma to source cosmic  magnetic fields through the 
    Harrison mechanism. We analyze the coupled set of perturbed Maxwell and Boltzmann 
    equations for a plasma in which the matter and radiation components exhibit relative 
    bulk motions at the background level. We find that, to first order in cosmological 
    perturbations, bulk flows with velocities compatible with current Planck limits 
    ($\beta<8.5\times 10^{-4}$ at $95\%$ CL) could  generate magnetic fields with an 
    amplitude $10^{-21}$ G  on 10 kpc comoving scales at the time of completed galaxy 
    formation which could be sufficient to seed a galactic dynamo mechanism.  
\end{abstract}
\maketitle

\paragraph{Introduction.}
	The origin of the magnetic fields with strengths in the range of the $\mu$G found in 
	galaxies and permeating the intergalactic medium in clusters is a long-standing 
	question in astrophysics and cosmology \cite{Widrow:2002ud}. Even more puzzling is the 
	presence of magnetic fields in voids with strengths $3\times 10^{-16}$ G as those 
	detected in \cite{Neronov:1900zz}. The evolution of primordially generated 
	magnetic fields from the early Universe to the onset of structure formation seems to be 
	well understood \cite{Banerjee:2004df, Durrer:2013pga, Subramanian:2015lua}, and
	there are compelling astrophysical mechanisms, i.e. dynamos, that can amplify a 
	preexisting magnetic field several orders of magnitude \cite{Davis:1999bt, 
	Widrow:2002ud}. However, a definite mechanism that can \emph{produce} the primordial seed 
	fields is still lacking.
	
	There are different proposed solutions, that can be classified as cosmological or
	astrophysical, addressing the origin of the primordial fields. In  the cosmological 
	mechanisms, magnetic fields are generated in the early Universe, typically during 
	inflation \cite{Turner:1987bw, Maroto:2000zu} or in the electroweak \cite{Vachaspati:1991nm} or 
	QCD \cite{Quashnock:1988vs}	phase transitions. On the other hand, in astrophysical 
	mechanisms, magnetic fields are generated by motions in the plasma during galaxy 
	formation. In general, the amplitude of the seeds generated by these mechanisms is 
	too small to explain the observed fields even with dynamo amplification. 
	Depending on the dynamo amplification rate, a seed field with a strength in the 
	range $10^{-23}-10^{-16}$ G at galaxy formation and coherent on comoving scales of 
	10 kpc  is required to reach the amplitude of the detected galactic fields 
	\cite{Davis:1999bt}.   
	
	Among the  astrophysical proposals, a particularly appealing one is the so-called 
	Harrison mechanism. In his pioneering work \cite{harrison1970generation}, 
	Harrison realized that vorticity in the photon-baryon plasma would lead to the 
	production of electromagnetic fields. The main obstacle \cite{1987QJRAS..28..197R}
	for the Harrison mechanism to work is to achieve vortical motions in the fluid. 
	Within $\Lambda$CDM, to first order in perturbation theory, vorticity and vector modes 
	decay so, even if they are initially large, only small magnetic fields can be 
	generated \cite{Ichiki:2011ah}. Different routes have been explored to overcome this difficulty.
	It is possible to source 
	vector modes, e.g. via topological defects, but it was shown in 
	\cite{Hollenstein:2007kg} that if vorticity is transferred	only by gravitational 
	interactions, it does not lead to production of magnetic fields. On the other hand, vorticity 
	and magnetic fields are indeed generated to second order in perturbation theory in 
	standard $\Lambda$CDM \cite{Takahashi:2005nd, Fenu:2010kh, Saga:2015bna}, but are 
	consequently very small.
	
	Recently, it has been shown that vorticity in the photon-baryon plasma can also be 
	produced if bulks flows of matter with respect to radiation are present 
	\cite{Cembranos:2019plq}. In such a case, first order scalar metric perturbations 
	induce non-decaying vortical motions in the different plasma components. 
	
	The existence of large-scale bulk flows in excess of $\Lambda$CDM predictions has been 
	a matter of debate in recent years. While some papers claim to find evidence of 
	unusually large flows \cite{Kashlinsky:2008ut,Atrio-Barandela:2014nda}, most of the 
	works find results consistent with $\Lambda$CDM \cite{Ade:2013opi, Scrimgeour:2015khj}. 
	In particular, the largest-scale limits to date on the amplitude of the bulk flow has 
	been set by Planck collaboration \cite{Ade:2013opi} from measurements of the kinetic 
	Sunyaev-Zeldovich effect in clusters and is given by $\beta<8.5\times 10^{-4}$ at 
	$95\%$ CL on 2 Gpc scales.
		
	In this work we find that even a small  background bulk velocity, compatible with the 
	Planck limit, is able to generate vorticity to source magnetic fields above the dynamo 
	threshold through the Harrison mechanism.\\

\paragraph{Plasma system.}
	Let us assume a homogeneous plasma system composed of photons, protons and electrons 
	with background bulk velocities $\v{\beta}_\gamma$, $\v\beta_p$ and $\v\beta_e$ respectively. 
	As shown in \cite{Cembranos:2019plq}, to first order in $\beta$ it is always possible 
	to find a center of mass frame in which the metric takes the Robertson-Walker (RW) 
	form. Thus, including scalar perturbations in the Newtonian gauge the metric reads
	\begin{align}
		\di s^2 = a^2(\tau)\Big\{&-(1+2\psi)\,\di\tau^2+(1-2\phi)\,\di\v{x}^2\Big\}\ ,
	\end{align}
	and the perturbed fluid velocities can be written as $\v v_s=\v\beta_s+\delta \v v_s$ 
	with $s=\gamma, e, p$. In the following we will work to first order in bulk velocities 
	and first order in scalar metric perturbations, ignoring the contribution of vector 
	and tensor modes which, as shown in \citep{Cembranos:2019plq}, would appear as
	$\Od{\beta^2}$ corrections.
	
	The behaviour of the electron-proton-photon plasma is described by a set of coupled
	Boltzmann equations which, in a locally inertial frame ($\di t\equiv a(1+\psi)\di\tau$),  
	reads \cite{Cembranos:2019plq}
	\begin{subequations}
	\begin{align}
		\frac{Df_\gamma}{\di t} &= \mathcal{C}_{\gamma e}[f_\gamma] 
				+ \mathcal{C}_{\gamma p}[f_\gamma]\ ,\\
		\frac{Df_e}{\di t} &= \mathcal{C}_{e\gamma}[f_e] + \mathcal{C}_{ep}[f_e]\ ,\\
		\frac{Df_p}{\di t} &= \mathcal{C}_{p\gamma}[f_p] + \mathcal{C}_{pe}[f_p]\ ,
	\end{align}
	\end{subequations}
	where the collision terms take into account both Thomson scattering and the Coulomb
	interaction between electrons and protons. The evolution of the momentum of the fluids
	can be followed performing the appropiate integrals over the phase-space distributions.
	Expressing the results in conformal time $\tau$, integrating over the comoving 
	momentum $q^i$, and defining
	\begin{equation}\label{eq:def_DQ}
		\frac{DQ^i_s}{\di\tau} \equiv 2a^{-4}\int\frac{\di^3q}{(2\pi)^3}q^i\frac{Df_s}{\di\tau}\ ,
			\qquad s=\gamma,e,p\ .
	\end{equation}
	we have
	\begin{subequations}
	\begin{align}
		\frac{D Q^i_\gamma}{\di\tau} &= C^i_{\gamma e} + C^i_{\gamma p}\ ,
			\label{eq:Q_photons}\\
		\frac{D Q^i_e}{\di\tau} &= C^i_{e\gamma} + C^i_{ep}\ ,
			\label{eq:Q_electrons}\\
		\frac{D Q^i_p}{\di\tau} &= C^i_{p\gamma} + C^i_{pe}\ .
			\label{eq:Q_protons}
	\end{align}
	\end{subequations}
	Additionally, from momentum conservation in Coulomb and Thomson scattering we have
	$C^i_{s_1s_2}=-C^i_{s_2s_1}$. The electron coupling due to Thomson scattering is 
	\cite{Cembranos:2019plq}
	\begin{align}
		C^i_{\gamma e} = \frac{4}{3}\rho_\gamma an_e\sigma_T \Big(&\Delta\beta_{\gamma e}^i
			+ \Delta v_{\gamma e}^i + \beta^i_\gamma\delta_{n_e}-\beta_e^i\delta_\gamma \nonumber\\
			&-\frac{3}{4}\beta_{e\,j}\pi^{ij}_\gamma +\Delta\beta^i_{\gamma e}\psi\Big)\ ,
	\end{align}
	where $\delta_{n_e}=\delta n_e/n_e$ is the perturbation of the number of free electrons
	and $\pi^{ij}_\gamma$ is the photon shear tensor.
	The corresponding Thomson coupling between protons and photons can be obtained with
	the substitution $e\to p$ and $\sigma_T\to (m_e/m_p)^2\sigma_T$. The coupling due to
	Coulomb scattering takes a similar form \cite{Fenu:2010kh}
	\begin{align}
		C^i_{e p} = -e^2an_pn_e\eta_C\Big(&\Delta\beta_{ep}^i + \Delta v^i_{ep} 
				+\Delta\beta^i_{ep}\delta_{n_e}\nonumber\\
			&-\beta_e^i\Delta n_{ep}+\Delta\beta_{ep}^i\psi\Big)\ ,
	\end{align}
	where $\eta_C$ is the electrical resistivity and we have defined, 
	for two species $a$ and $b$, the following quantities
	\begin{equation}
		\Delta n_{ab}\equiv \delta_{n_a} - \delta_{n_b},\quad
		\Delta \beta^i_{ab}\equiv \beta_a^i - \beta_b^i,\quad
		\Delta v^i_{ab}\equiv \delta v_a^i - \delta v_b^i .
	\end{equation}
	The left-hand side of the Boltzmann equation \eqref{eq:def_DQ} can be splitted into
	the usual geodesic evolution plus a term taking into account the presence of 
	macroscopic electromagnetic fields. We define the electric  and magnetic components
	of the electromagnetic strength $F_{\mu\nu}$ in the perturbed RW metric as 
	$\mathcal{E}_i=(1+\phi)F_{i0}$ and $\mathcal{B}_i=\frac{1}{2}\varepsilon^{ijk}F_{jk}$. 
	These fields affect the motion of charged particles through the Lorentz force which 
	takes the standard form
	\begin{equation}
		\left(\d{q_i}{\tau}\right)_\text{EM} = e\left(\mathcal{E}_i 
			+ \varepsilon_{ijk}\frac{q^j}{\epsilon}\mathcal{B}^k\right)\ .
	\end{equation}
	where $\epsilon \equiv \sqrt{m^2a^2+q^2}$ is the comoving energy.
	Notice that, in the absence of bulk flows,  scalar perturbations cannot generate 
	magnetic fields to first order in perturbation theory. Therefore, in our scenario, 
	$\mathcal{B}^i$ can only arise as a cross-product of $\beta^i$ with perturbations. The 
	electric field, on the other hand, can be splitted into a homogeneous piece of 
	$\Od{\beta}$ and a perturbation, $\mathcal{E}^i = \mathcal{E}^i_{(\beta)} 
	+ \delta \mathcal{E}^i$. Adding the electromagnetic force to \eqref{eq:Q_electrons}, 
	the evolution of the velocity of the electrons is
	\begin{align}
		m_e&n_e\Big\{\left(\partial_\tau+\alpha+\mathcal{H}\right)\left(\beta_e^i+\delta v_e^i\right)
				+ \left(\beta^i_e\delta^j_k+\beta_e^j\delta_k^i\right)\partial_j\delta v_e^k\nonumber\\
			&\qquad +\partial^i\psi - 4\beta^i_e\dot{\phi}
				+\frac{e}{m_ea}(1+\delta_{n_e})\mathcal{E}^i\Big\}\nonumber\\
			&= C^i_{e\gamma}+C^i_{ep}\ .
	\end{align}
	The first line contains, in addition to the usual Hubble dilution term, a coefficient 
	$\alpha=\partial_\tau(a^3n_e)/(a^3n_e)$ representing a possible variation in the 
	comoving number of free electrons at the background level, e.g. due to recombination, and
	the effective shear stress induced by the bulk motion of the fluid 
	$\pi_{ij}\sim \beta_i\delta v_j$. The second line contains the effect of metric 
	perturbations, both the standard one and the correction induced by the presence
	of cosmological bulk flows \cite{Cembranos:2019plq}. The metric contribution is
	irrelevant for the Harrison mechanism but it will be important to study the evolution
	of the photon-baryon plasma vorticity.
	Finally, the last term takes into account the electromagnetic effects. A 
	similar result can be found for	protons after changing the relevant subscripts and the 
	electric charge $e\to -e$. Subtracting the equations for electrons and protons, we 
	obtain an expression for the velocity difference
	\begin{align}\label{eq:Delta_vep_evolution}
		&\left(\partial_\tau+\alpha+\mathcal{H}\right)\left(\Delta\beta_{ep}^i+\Delta v_{ep}^i\right)\nonumber\\
			&\quad\qquad+ \left(\beta_e^i\theta_e+\beta_e^j\partial_j\delta v_e^i-(e\leftrightarrow p)\right)
				- 4\Delta\beta^i_{ep}\dot{\phi}\nonumber\\
			&\quad\qquad+\frac{e}{m_ea}\left(\mathcal{E}^i_{(\beta)}+\delta \mathcal{E}^i 
				+ \delta_{n_e}\mathcal{E}^i_{(\beta)}\right)\nonumber\\			
		&\quad = \frac{1}{m_en_e}\left(C_{e\gamma}^i + C^i_{ep}\right)\ ,
	\end{align}
	where we have used the fact that $m_p\gg m_e$. Below we show how this expression, combined
	with the Maxwell equations, gives rise to magnetic fields.\\
	
\begin{figure}[t]
		\includegraphics[scale=0.5]{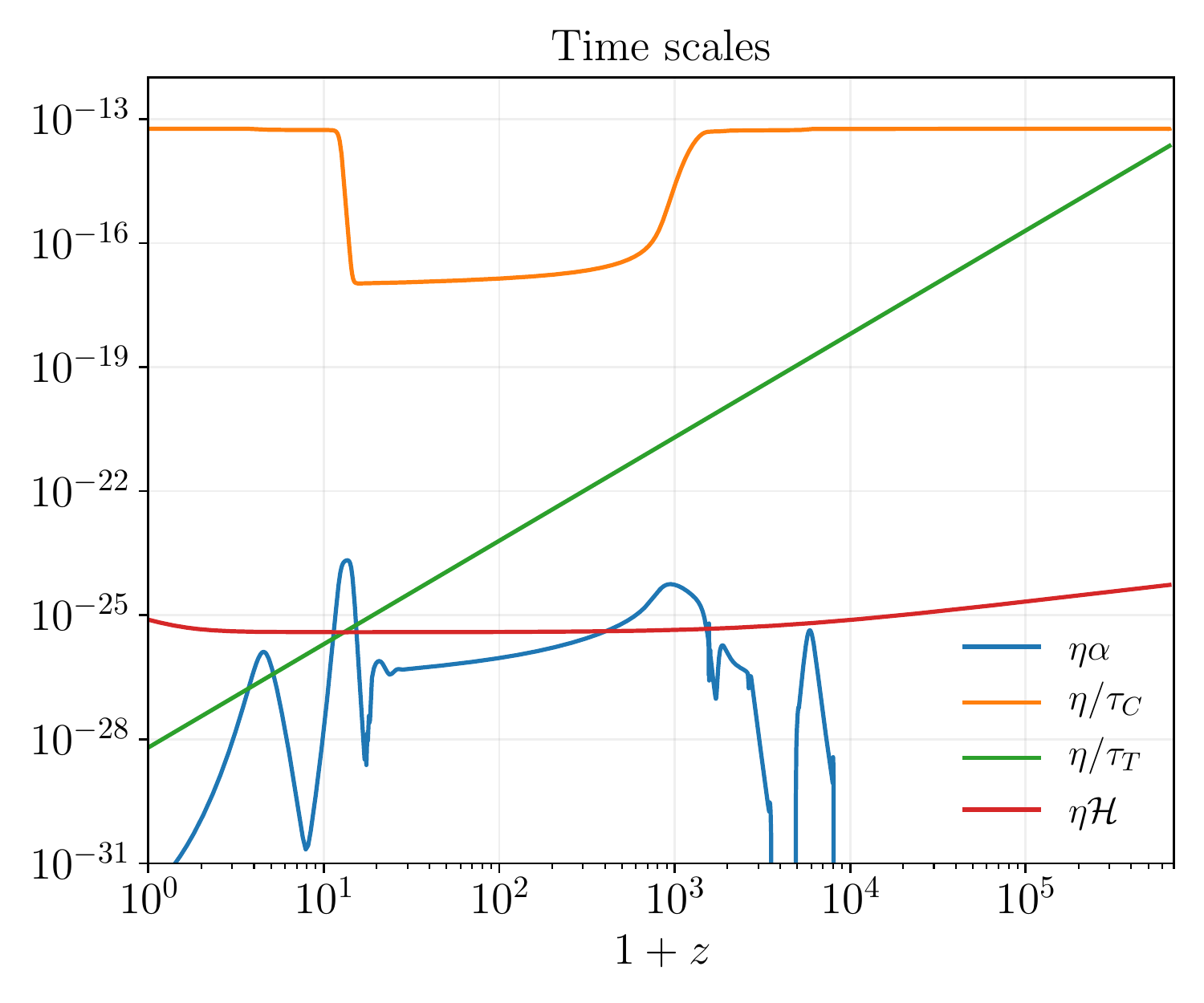}
		\caption{Ratios of the relevant scales of the problem, with respect to the
		dominant one: the electrical resistivity $\eta$. During the period of interest
		the next scale in the hierarchy is the Coulomb time scale. Early enough in
		time, Thomson scattering becomes more efficient than Coulomb scattering.}
		\label{fig:hierarchy}
	\end{figure}

\paragraph{Time scales.}
	Following \cite{Fenu:2010kh} we define the time scales relevant for the
	system \eqref{eq:Delta_vep_evolution}, assuming a matter-dominated universe.
	\begin{itemize}
		\item Electrical resistivity.
			\begin{equation}
				\eta\equiv\frac{\eta_C}{a}\simeq\frac{10\pi e^2\sqrt{m_e}}{aT^{3/2}}
					\simeq 10^{-9}\ \text{s}\left(\frac{1+z}{10^3}\right)^{-1/2}.
			\end{equation}
		\item Coulomb time scale.
			\begin{equation}
				\tau_C \equiv \frac{m_e}{ae^2n_e\eta_C}\simeq 
				\frac{2\times 10^4\ \text{s}}{x_e}\left(\frac{1+z}{10^3}\right)^{-1/2}.
			\end{equation}
		\item Thomson time scale.
			\begin{equation}
				\tau_T \equiv \frac{m_e}{a\sigma_T\rho_\gamma}\simeq 5\times 10^{11}\ \text{s}
					\left(\frac{1+z}{10^3}\right)^{-3}.
			\end{equation}
	\end{itemize}
	There are other time scales in the problem like the cosmological ones, $\mathcal{H}^{-1}$
	and $k^{-1}\simeq 10^{14}\ \text{s}\ (\text{Mpc}^{-1}/k)$, and the time scale of
	recombination $\alpha = \dot{x}_e/x_e$. The ratio of these scales with respect
	to $\eta$ is represented in Fig. \ref{fig:hierarchy}.
	
	There is a very strong hierarchy of scales, with 
	$\eta\ll\tau_C\ll\tau_T,\mathcal{H}^{-1},\alpha^{-1}$. In the next section, we will
	use this fact to find an approximate solution of the system.\\
	
\paragraph{Production mechanism.}
	The main physical mechanisms at work can be nicely illustrated analyzing the behaviour
	of the bulk velocities. The relevance of the previous time scales will be made 
	explicit if we write the equations in terms of $e^i_{(\beta)}$, where 
	$\mathcal{E}_{(\beta)}^i=a^{5/2}n_e\tau_C\,e^i_{(\beta)}$. At the background level, 
	the leading $\Od{\beta}$ piece of \eqref{eq:Delta_vep_evolution}, plus the relevant 
	Maxwell equation, yields
	\begin{subequations}\label{eq:system_beta}
	\begin{align}
		\Delta\dot{\beta}_{ep}^i 
			+ \left(\frac{1}{\tau_c}+\alpha+\mathcal{H}\right)\Delta\beta^i_{ep}
			+ \frac{1}{a^{1/2}\eta}e^i_{(\beta)} &= \mathcal{T}^i_\beta\ ,\\
		\dot{e}^i_{(\beta)} - \frac{a^{1/2}}{\tau_c}\Delta\beta_{ep}^i &= 0\ ,
	\end{align}
	\end{subequations}
	where the Thomson dragging term is $\mathcal{T}^i_\beta\equiv \frac{4}{3\tau_T}\Delta\beta^i_{\gamma e}$.
	The result is a very simple dynamical system where, as discussed in the previous 
	section, the strong hierarchy of scales present in the problem
	allows us to simplify the analysis keeping only the leading $\Od{\eta}$ behaviour.
	The homogeneous part of this system (without the source) corresponds to the usual electron-proton plasma (without photons).
	If the system is placed out of the equilibrium $\Delta\beta_{ep}^i=e_{(\beta)}^i=0$
	configuration, an electric field is created in response, acting as a restoring force. 
	The homogeneous solutions oscillate with characteristic frequency $\omega\simeq 1/\sqrt{\eta\,\tau_C}$
	and are damped with a damping coefficient $\Gamma \simeq 1/2\tau_C$. The presence
	of photons modifies this picture. Due to the large mass difference, $m_p\gg m_e$, the
	Thomson coupling of photons to electrons is much more effective than to protons, 
	producing a differential dragging and introducing the source $\mathcal{T}^i_\beta$. The
	particular solution of the system \eqref{eq:system_beta} can be found to be
	\begin{subequations}
	\begin{align}
		\Delta \beta^i_{ep} &= \eta\,\tau_C\,\dot{\mathcal{T}}^i_\beta + \Od{\eta^2}\ ,\\
		e^i_{(\beta)} &= a^{1/2}\,\eta\,\mathcal{T}^i_\beta + \Od{\eta^2}\ .
	\end{align}
	\end{subequations}
	This is the essence of the Harrison mechanism: the Thomson dragging of the photons
	produces an electric field proportional to the photon-baryon velocity difference.
	Notice that a homogeneous electric field is generated, pointing in the bulk flow
	direction and with a small amplitude $\mathcal{E}_{(\beta)}\lesssim 10^{-30}\ \text{G}(1+z)^2$,
	according to the current Planck limits for $\beta$.
	The same kind of analysis can be carried out to prove that 
	$\Delta n_{ep},\Delta v_{ep}^i = \Od{\eta\,\tau_C}$ and from 
	\eqref{eq:Delta_vep_evolution} we get the leading order result
	\begin{equation}
		\delta \mathcal{E}^i = \frac{a}{en_e}C^i_{e\gamma} - \delta_{n_e}\mathcal{E}^i_{(\beta)} + \Od{\eta}\ . \label{deltaE}
	\end{equation}
	 In Fourier space, we decompose the velocity and the 
	electromagnetic fields into vortical and longitudinal components as
	\begin{subequations}
	\begin{align}
		\delta \v{v}_s &= \chi_s\left(\hat{\beta}-(\hat{\beta}\cdot\hat{k})\,\hat{k}\right) 
			-\frac{\ii}{k}\theta_s\,\hat{k}\ ,\\
		\v{\mathcal{E}} &= \mathcal{E}^\bot\left(\hat{\beta}-(\hat{\beta}\cdot\hat{k})\,\hat{k}\right)
			+\mathcal{E}^\parallel\,\hat{k}\ ,\\
		\v{\mathcal{B}} &= \ii\, \mathcal{B}\left(\hat{\beta}\wedge\hat{k}\right)\label{eq:mag_ang_struct}\ .
	\end{align}
	\end{subequations}
	From the Maxwell equations, including perturbations, we have
	\begin{equation}
		\dot{\mathcal{B}} = -k\,\delta\mathcal{E}^\bot + k\phi\,\mathcal{E}_{(\beta)}^\bot\ . \label{mag}
	\end{equation}
	Plugging in the expression obtained for the electric field \eqref{deltaE} and written in terms
	of the physical magnetic field $\v{B}\equiv a^{-2}\v{\mathcal{B}}$, that can be obtained
	projecting with the tetrad of a locally inertial observer \cite{Durrer:2013pga}, Eq. \eqref{mag} reads
	\begin{align}\label{eq:final_evolution}
		\d{}{\tau}\left(a^2B\right) = -\frac{4a^2k\sigma_T\rho_\gamma}{3e}
			\Big(&\Delta\chi_{\gamma e} + \beta_e(\delta_{n_e}-\delta_\gamma)\nonumber\\
			&+\Delta\beta_{\gamma e}(\psi-\phi)\Big)\ .
	\end{align}
	This is the final equation governing the production of magnetic fields.	It generalizes the 
	Harrison mechanism to the case in which there are bulk flows in the plasma. It is also 
	analogous to the one obtained in previous studies of production	of magnetic fields in 
	second order cosmological perturbation theory \cite{Fenu:2010kh,Saga:2015bna}. Details 
	on the evolution of the cosmological bulk flows $\beta$, and the vorticity produced by 
	these flows can be found in \cite{Cembranos:2019plq}.
	 \\
	
\paragraph{Evolution and results.}	
	The magnetic field power spectrum is defined by
	\begin{equation}
		\av{B_i(z,\v{k})B_j^*(z,\v{k}')} = \delta (\v{k}-\v{k}')(\hat{\beta}\wedge\hat{k})_i
			(\hat{\beta}\wedge\hat{k})_jP_B(z, k)\,,
	\end{equation}
	as
	\begin{equation}
		P_B(z,k) = |T_B(z,k)|^2\frac{2\pi^2}{k^3}\mathcal{P}_\mathcal{R}(k)\ ,
	\end{equation}
	where $\mathcal{P}_\mathcal{R}(k)$ is the usual nearly scale-invariant primordial curvature power spectrum and
	$T_B(z,k)$ is the magnetic field transfer function computed using 
	\eqref{eq:final_evolution}. In Figs. 2 and 3 the comoving magnetic field 
	$(1+z)^{-2}|T_B|\mathcal{P}_\mathcal{R}^{1/2}$ is plotted as a function of redshift and scale 
	respectively.
	
	There are two points worth emphasizing. On the one hand, the magnetic power spectrum on 
	small and large scales has a power-law behaviour 
	\begin{equation}
		\sqrt{k^3P_B(z<100, k)}\propto\left\{\begin{array}{l}
			k^{1.2}\ ,\quad k\gg 0.1\ \text{Mpc}^{-1},\\[.9mm]
			k^{2.8}\ ,\quad k\ll 0.1\ \text{Mpc}^{-1},
		\end{array}\right.
	\end{equation}
	so that the magnetic field is steeply rising as $k^{1.2}$ on small scales, until the 
	turbulence scale kicks in. On the other hand, the comoving magnetic field is 
	continuously produced, with an important boost  at recombination and remaining 
	essentially constant for $z<100$.
		
	Following \cite{Fenu:2010kh}, we also define the magnetic 
	field smoothed over a comoving scale $L$ as
	\begin{equation}
		B_L^2(z) = \frac{1}{2\pi^2}\int^\infty_0 \di k\, k^2P_B(z, k)\exp\left(-\frac{k^2L^2}{2}\right)\ .
	\end{equation}
		
	\begin{figure}[t]
		\includegraphics[scale=0.5]{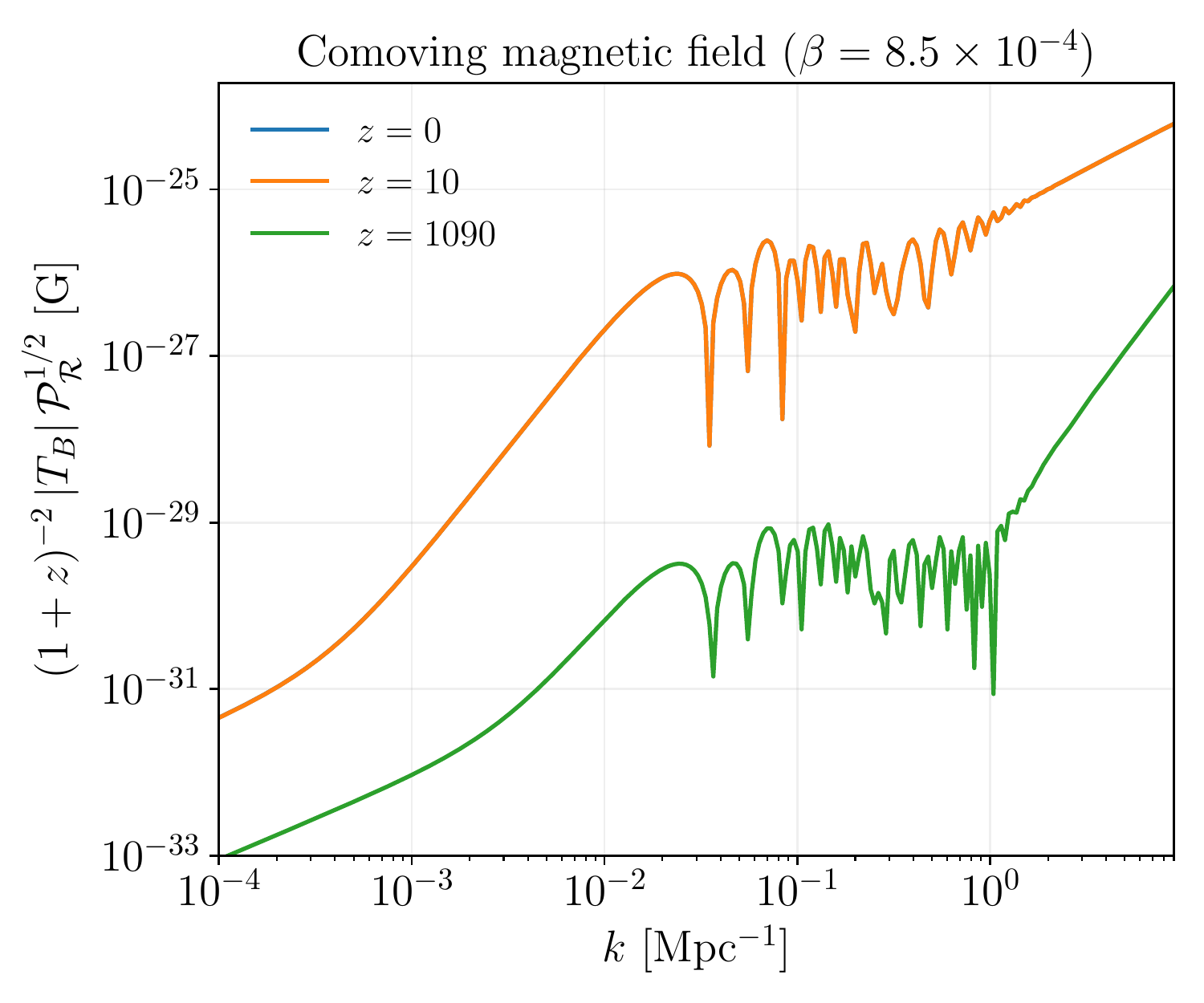}
		\caption{Comoving magnetic field as a function of the scale for different redshifts.
		Notice that the $z=0$ and $z=10$ curves overlap. Even though there is an important production immediately after 
		decoupling, afterwards the comoving magnetic field is constant at all scales and 
		it is not affected by reionization.}
		\label{fig:evolutionk}
	\end{figure}
	
	\begin{figure}[t]
		\includegraphics[scale=0.5]{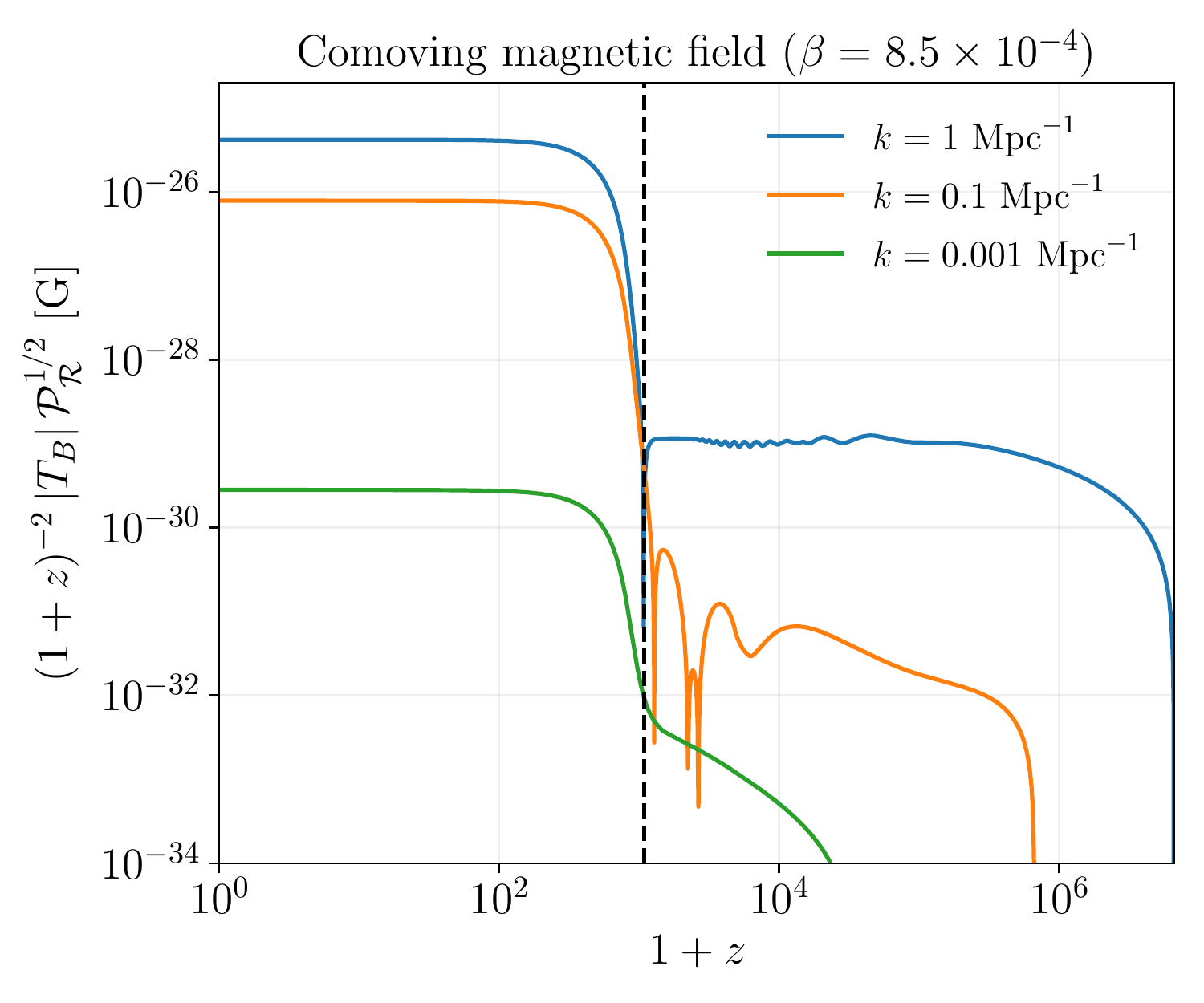}
		\caption{Comoving magnetic field as a function of the redshift for different
		scales. The magnetic field presents some features inherited from the acoustic
		oscillations before decoupling. The main production takes place during and 
		immediately after decoupling. Once the photon-baryon plasma is decoupled, the
		comoving magnetic field is constant.}
		\label{fig:evolutionz}
	\end{figure}
	
	\begin{figure}[t]
		\includegraphics[scale=0.5]{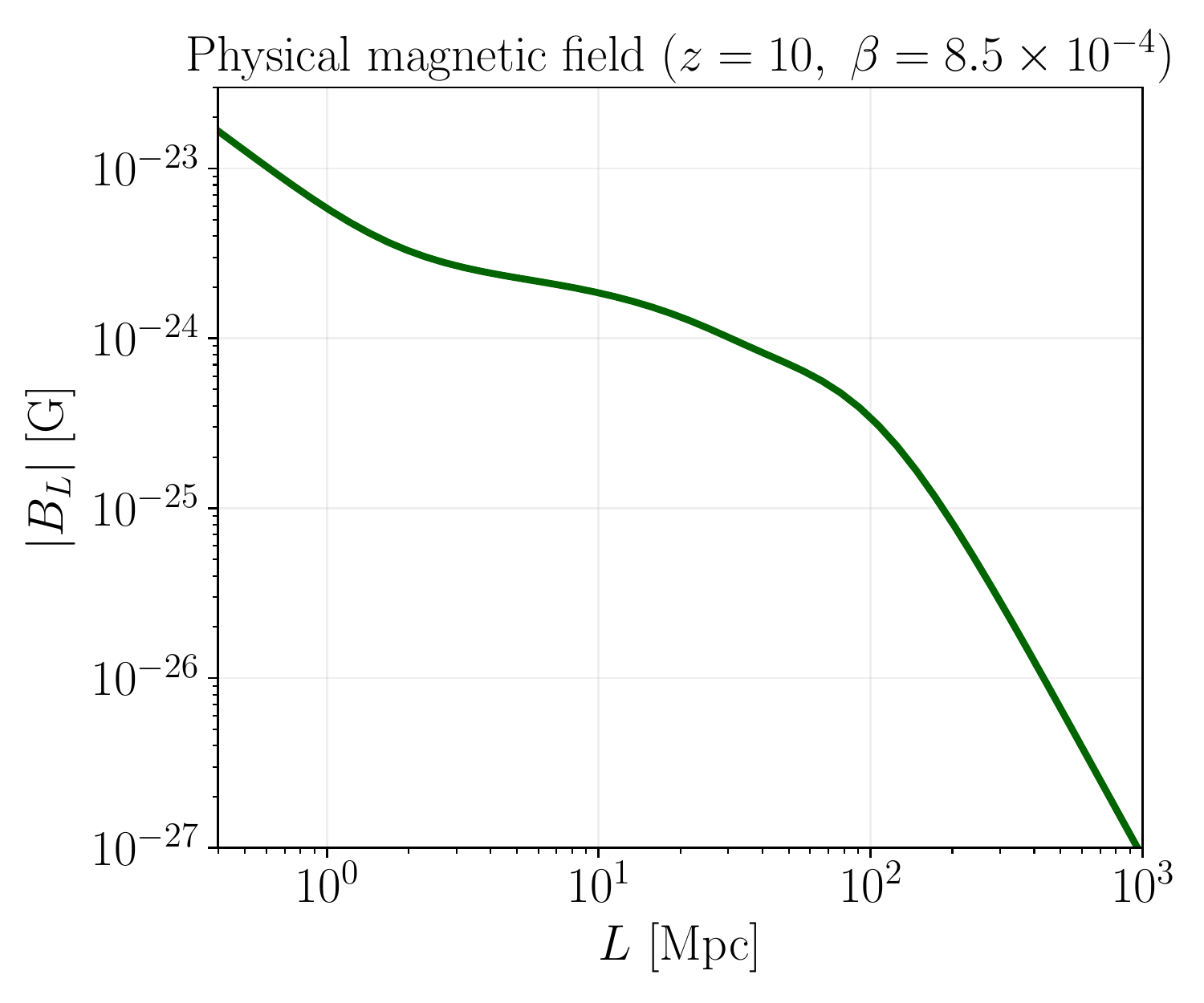}
		\caption{Physical magnetic field smoothed over a given scale $L$. It is evaluated
		at a redshift $z=10$, where the dynamo mechanism should begin to operate 
		\cite{Widrow:2002ud}. 
		Since the comoving field is constant at late times, the results can be easily rescaled to any
		redshift.}
		\label{fig:physical_magnetic}
	\end{figure}
	The magnetic field $B_L$ at the time of galaxy formation $z_\text{gf}=10$ is depicted 
	in Fig.  \ref{fig:physical_magnetic}. The numerical computation of the transfer 
	function becomes harder for smaller scales, and	some of the usual approximations in 
	CMB calculations cannot be trusted for scales $k > 10\ \text{Mpc}^{-1}$ 
	\cite{Blas:2011rf}. Therefore, we only compute the spectrum up to scales 
	$k = 9\ \text{Mpc}^{-1}$. The field $B_L$ can be well approximated as a power law 
	at small scales, yielding the approximate result
	\begin{align}
		|B_L(z<100)| &\simeq 5.7\times 10^{-24}\ \text{G}\ \left(\frac{L}{\text{Mpc}}\right)^{-1.2}\nonumber\\
		&\qquad \times\left(\frac{1+z}{11}\right)^{2}
			\left(\frac{\beta}{8.5\times 10^{-4}}\right)\ ,
	\end{align}
	for $L< 1\ \text{Mpc}$ where $\beta$ is the relative bulk velocity between photons
	and baryons. These results show that,	
	although the field seems too weak to directly account for the intergalactic magnetic 
	fields or magnetic fields in voids, the mechanism proposed provides a 
	seed field large enough to potentially explain the galactic magnetic fields, after
	a suitable dynamo amplification.\\
	
\paragraph{\textbf{Acknowledgements.}}
	This work has been supported by the MINECO (Spain) project FIS2016-78859-P(AEI/FEDER,
	UE).

\bibliography{MagneticBiblio}

\end{document}